\documentclass[prl,aps,preprint,showpacs]{revtex4}

\usepackage{graphicx}

\begin{document}

  \title{Alloy Stabilized Wurtzite Ground State Structures of
    Zinc-Blende Semiconducting Compounds}

  \author{H. J. Xiang}
  \affiliation{National Renewable Energy Laboratory, Golden, Colorado
    80401, USA}

  \author{Su-Huai Wei}

  \affiliation{National Renewable Energy Laboratory, Golden, Colorado 80401, USA}

   \author{Shiyou Chen}
  
   \affiliation{Surface Science Laboratory and Department of Physics,
     Fudan University, Shanghai 200433, China}

   \author{X. G. Gong}
   \affiliation{Surface Science Laboratory and Department of Physics,
     Fudan University, Shanghai 200433, China}

  \date{\today}

  \begin{abstract}    
    The ground state structures of the A$_x$B$_{1-x}$C wurtzite (WZ) alloys
    with $x=$0.25, 0.5, and 0.75 are revealed by a ground state search using the
    valence-force field model and density-functional theory total energy calculations.
    It is shown that the ground state WZ alloy always has a lower strain energy and
    formation  enthalpy  than the corresponding zinc-blende (ZB)
    alloy. Therefore, we propose that the WZ phase can be stabilized through 
    alloying. This novel idea is supported by the fact that the
    WZ AlP$_{0.5}$Sb$_{0.5}$, AlP$_{0.75}$Sb$_{0.25}$,
    ZnS$_{0.5}$Te$_{0.5}$, and ZnS$_{0.75}$Te$_{0.25}$ alloys in the lowest
    energy structures are more stable than the corresponding ZB alloys.
    To our best knowledge, this is the first example where the alloy
    adopts a structure distinct from both parent phases.
  \end{abstract}


  \maketitle
  
  III-V and II-VI semiconductors usually crystallize into one of two forms:
  hexagonal wurtzite (WZ) and cubic zinc blende (ZB) structures.
  The ZB and WZ structures have the same local tetrahedral environment
  and start to differ only in their third-nearest-neighbor atomic
  arrangement.
  Despite the structural similarity, there are some significant
  differences in the electronic and optical properties \cite{Yeh1994,Schlfgaarde1997}. 
  Compared to the hexagonal structure, the cubic phase has a more
  isotropic property, higher carrier mobility, lower phonon scattering,
  and often better doping efficiency. In contrast, the WZ phase has
  a larger band gap (usually direct), a spontaneous electric
  polarization, and a lower propagating speed of dislocations and thus
  an improved lifetime of the laser diodes \cite{Sugiura1997}. 
  For certain device applications, one phase is preferred over the
  other. To have a controllable way to synthesize the desired phase, 
  it is important to understand the mechanism for stabilizing a
  certain structure.
  
  In general, the WZ structure is preferred over the ZB
  structure when the ionicity of a compound is high \cite{Garcia1993}. This is 
  because the ideal WZ structure has a larger
  Coulomb interaction energy with a larger Madelung constant, whereas  
  the ZB structure leads to a better covalent bond formation
  \cite{Phillips1973,John1974,Chelikowsky1978,Yeh1992}.
  To change the stability, one often grows materials into different forms. 
  For example, many ZB compounds can adopt the hexagonal WZ
  structure when forming nanowires (NWs)
  \cite{Koguchi1992,Shan2006,Patriarche2008}. 
  Empirical calculations suggested that the stability of the WZ NW is due to the fact
  that the WZ NW has less surface atoms than the ZB NW with a similar
  diameter \cite{Akiyama2006,Dubrovskii2008}. 
  Theoretical calculations also showed that
  stability of WZ compounds such as GaN can be changed when carriers are 
  introduced through doping \cite{Dalpian2004,Dalpian2006}.
  Moreover, metastable phases can be synthesized by employing non-equilibrium
  growth techniques. For example, metastable ZB GaN can be grown on
  cubic substrates \cite{Lazarov2005}.  
    
  In this paper, we show for the first time that the ground state (GS) WZ alloy (WZA) always 
  has a lower strain energy than the corresponding ZB alloy (ZBA). Therefore,
  if strain energy is dominant in alloy formation, stable GS ternary WZAs 
  can form even though the binary constituents are more stable in the ZB phase.
  This provides an opportunity to form desired WZAs through alloying. Our first principles
  calculations confirm this idea, showing that WZ
  AlP$_{0.5}$Sb$_{0.5}$, AlP$_{0.75}$Sb$_{0.25}$, 
  ZnS$_{0.5}$Te$_{0.5}$, and ZnS$_{0.75}$Te$_{0.25}$ have lower total 
  energies than the ZB counterparts.   
  
\begin{figure}
    \includegraphics[width=7.5cm]{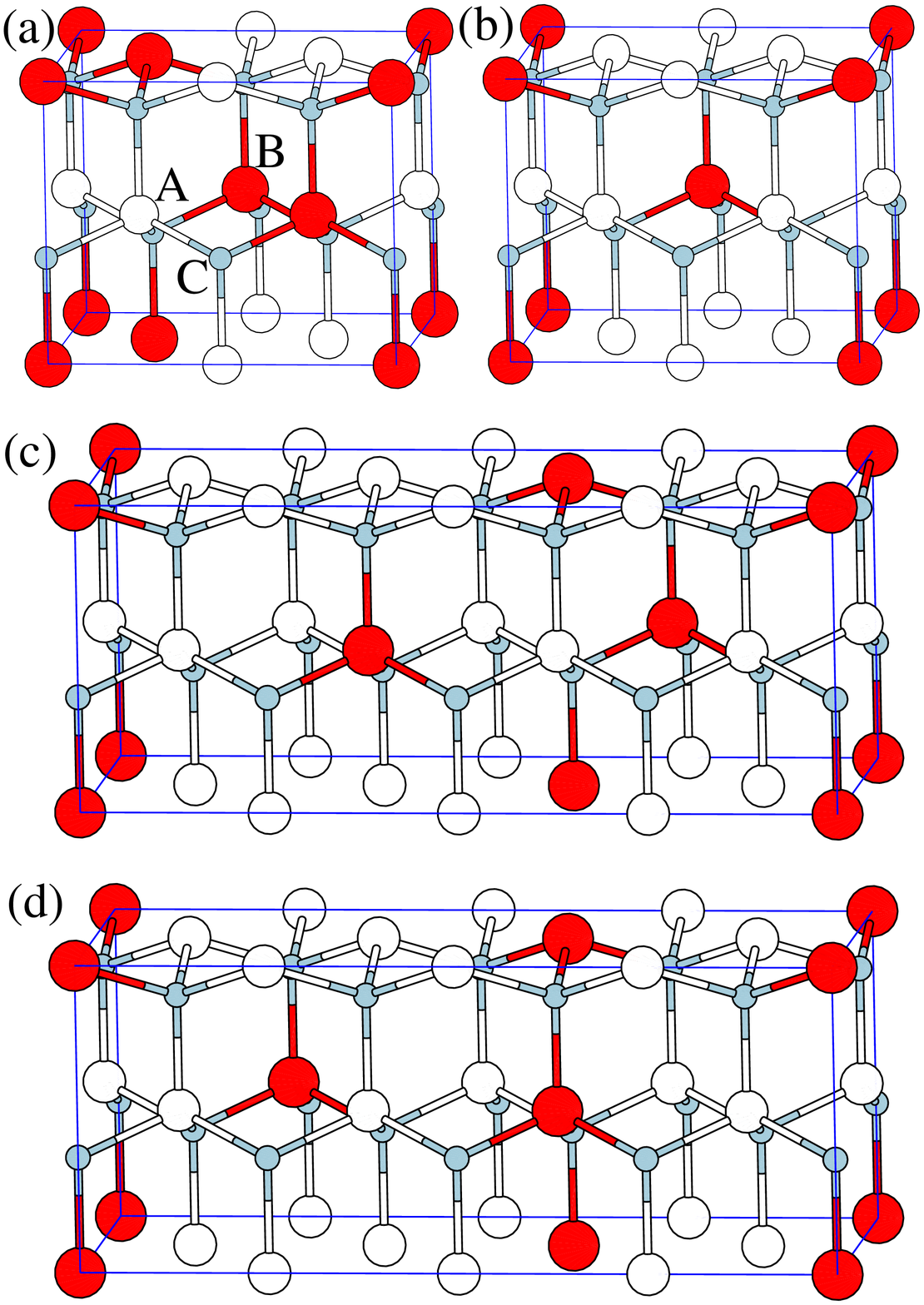}
    \caption{(a) The GS $Pna2_1$ structure of the WZ
      A$_{0.5}$B$_{0.5}$C alloy.
      (b) The $Pmn2_1$ structure, which is the lowest strain energy
      structure of the WZ A$_{0.25}$B$_{0.75}$C alloy. (c) The
      lowest strain energy structure ($P2_1$-I) of the
      WZ A$_{0.75}$B$_{0.25}$C alloy. 
      (d) The low strain energy structure ($P2_1$-II) of the
      WZ A$_{0.75}$B$_{0.25}$C alloy.}
    \label{fig1}
  \end{figure}

  The GS structures of ZBAs have been
  extensively studied \cite{Wei1990,Ferreira1989,Lu1994,Liu2007,Chen2008}.  
  For instance, it was shown that the GS ZB A$_{0.5}$B$_{0.5}$C alloy
  (Without loss of generality, B ion
  is assumed to have a larger radius than A ion, and C could be anion or
  cation)
  adopts the tetragonal chalcopyrite structure (space
  group I$\bar 4$2d, No. 122) \cite{Wei1990}.   
  However, the knowledge of the GS structures of WZAs 
  remains incomplete. Our previous work \cite{Xiang2008} showed that the GS
  structure of the A$_{0.5}$B$_{0.5}$C WZA is of the $\beta-$NaFeO$_2$
  type with the space group $Pna2_1$ (No. 33) as shown in
  Fig.~\ref{fig1}(a).
  Here, in this work, we identify that the GS structures of A$_{0.25}$B$_{0.75}$C and
  A$_{0.75}$B$_{0.25}$C WZAs  have the structures shown in  Fig.~\ref{fig1}(c) or (d) with the space group $P2_1$ (No. 4).   

  The formation enthalpy of isovalent semiconductor alloys
  A$_x$B$_{1-x}$C is defined as
  \begin{equation}
    \Delta H_f = E(x)-[xE_{AC} + (1-x)E_{BC}],
    \label{eq1}
  \end{equation}
  where $E_{AC}$, $E_{BC}$, and  $E(x)$ are the total energies of
  bulk AC and BC, and the A$_x$B$_{1-x}$C alloy with the same crystal
  structure (WZ or ZB). 
  It is well known that for lattice-mismatched isovalent
  semiconductor alloys, the major contribution to the formation
  enthalpy is the strain energy. The strain energy ($E_s$) could be
  described well by the VFF model
  \cite{Keating1966,Martin1970,Martins1984},
  which considers the deviation of the nearest-neighbor
  bond lengths and bond angles from the ideal bulk values.
  Here, we consider all
  possible supercells with up to 32 atoms per unit cell. For each supercell, we
  consider all possible configurations of alloys with x=0.25 and 0.75. 
  The VFF model is used to relax the structure
  and predict the energy of the configuration.  We considered
  Ga$_{x}$In$_{1-x}$N, AlP$_{x}$Sb$_{1-x}$,
  ZnS$_{x}$Te$_{1-x}$, and GaP$_{x}$As$_{1-x}$. They have various
  degrees of lattice mismatch: 10.1\%, 11.5\%, 11.8\%, and 3.7\%.
  Our calculations reveal
  the Lazarevicite structure (space group $Pmn2_1$, No. 31) shown in
  Fig.~\ref{fig1}(b) has the lowest strain energy for 
  A$_{0.25}$B$_{0.75}$C for all four different sets of VFF parameters
  \cite{Martins1984,Kim1996}.  The $Pmn2_1$ A$_{0.25}$B$_{0.75}$C
  structure has the same supercell as the $Pna2_1$
  A$_{0.5}$B$_{0.5}$C structure. One can get the $Pmn2_1$
  A$_{0.25}$B$_{0.75}$C
  structure by replacing one half of the A atoms in the $Pna2_1$
  A$_{0.5}$B$_{0.5}$C structure with B atoms so that each C atom has
  one neighbor A atom and three neighbor B atoms. The $Pmn2_1$ WZ
  A$_{0.25}$B$_{0.75}$C structure is similar to the famatinite ZB
  A$_{0.25}$B$_{0.75}$C structure \cite{Liu2007} in that they both
  have similar local environment for C atoms.

  For the WZ A$_{0.75}$B$_{0.25}$C alloy, we identify two low strain energy
  structures with the $P2_1$ space group (No. 4) [$P2_1$-I:
  Fig.~\ref{fig1}(c), and $P2_1$-II: Fig.~\ref{fig1}(d)].
  In contrast to   
  the $Pmn2_1$ A$_{0.75}$B$_{0.25}$C structure,
  there are some C atoms which have four A neighbor atoms in both
  structures.  In this sense, the $P2_1$-I and $P2_1$-II WZ
  A$_{0.75}$B$_{0.25}$C structures are similar to the Q8 and Q16 ZB
  A$_{0.75}$B$_{0.25}$C structures \cite{Wei1990,Lu1994}. 
  As shown in
  Table~\ref{table1}, the $P2_1$-I structure has
  the lowest strain energy. However, the strain energy difference between the
  $P2_1$-II and $P2_1$-I structures is very small, less than 0.3 meV/atom.
      
  To see if the GS structures predicted by VFF strain energy calculations are consistent to the 
  density functional theory (DFT) total energy calculations,
  we performed DFT calculations \cite{DFT,PAW,VASP,LDA} on the WZ 
  A$_{0.25}$B$_{0.75}$C and A$_{0.75}$B$_{0.25}$C alloys with the
  $Pmn2_1$,  $P2_1$-I, and $P2_1$-II structures. Our results are shown
  in Table~\ref{table1}. We can see that the $Pmn2_1$ structure is
  not the GS of the WZ A$_{0.25}$B$_{0.75}$C alloy because the
  $P2_1$ structures have a slightly lower total energy, even though the
  $Pmn2_1$ structure has a lower strain energy. 
  This can be explained in terms
  of the Coulomb interaction. For the A${_x}$B$_{1-x}$C alloy, the
  charge of A ions is different from that of B ions due to the
  different electronegativity. In this case, the Coulomb
  interaction is found to stablize the $P2_1$
  structures over the $Pmn2_1$ structure because the $P2_1$ structures 
  has larger charge fluctuation \cite{Magri1990}. Similar situation
  also occurs in ZBAs \cite{Chen2008}.

  After knowing the GS structures, we now compare the strain energy of the ZBA 
  and WZA  using the VFF model.  
  Our results are shown in Table~\ref{table2}. We can see that for all
  considered systems (Ga$_{x}$In$_{1-x}$N, AlP$_{x}$Sb$_{1-x}$,
  ZnS$_{x}$Te$_{1-x}$, and GaP$_{x}$As$_{1-x}$ with $x=0.25$, $0.5$,
  and $0.75$), the GS WZAs always have a lower strain energy
  than the GS ZBAs. The difference in the
  formation enthalpy mainly depends on the size of the lattice mismatch of
  alloy: For the first three A$_{x}$B$_{1-x}$C alloys with large lattice mismatch ($\Delta a > 10\%$), 
the strain energy difference $dE_s$ at $x=0.5$ is around 5 meV/atom,
  whereas, the difference $dE_s$ for GaP$_{0.5}$As$_{0.5}$ ($\Delta a < 4\%$), is only 0.7
  meV/atom.    
  
  Our above VFF calculations show that the WZ structure has a
  better ability to accomodate the strain in a lattice mismatched
  alloy than the ZB structure. This is due to the fact that the WZ
  structure has a larger degree of freedom to release the
  strain. First, for the binary 
  compound, the four-atoms unit-cell WZ structure has three free parameters
  ($a$, $c$, $u$). In contrast, the two-atoms unit-cell ZB structure only has one
  free parameter ($a$). Second, the WZA is also more flexible
  than the ZBA. As an example, we compare the 16-atoms
  WZ Pna2$_1$ and  8-atoms ZB chalcopyrite structures. In
  both structures, each C 
  atom bonds with two A and two B atoms. 
  In the ZB A$_{0.5}$B$_{0.5}$C chalcopyrite structure, 
  there are three free parameters.   
  However, there are fifteen free parameters in the WZ Pna2$_1$
  structure. The larger number
  of degree of freedom in the WZ Pna2$_1$ structure leads to an
  enhanced flexibility in strain relaxation. 
  
  For a better understanding of the strain relaxation in WZAs, we
  can also decompose the total strain energy into the contributions from
  each atom \cite{decompose}. In this way, we can tell which kind
  of atoms are mainly responsible for the different behavior between
  the WZ and ZB alloys. 
  This analysis shows that the main difference comes from the B ions with a
  large size. For example, the total contributions to the strain
  energy in the chalcopyrite (Pna2$_1$) AlP$_{0.5}$Sb$_{0.5}$ alloy
  (here A$=$P, B$=$Sb, and C$=$Al)
  from Al, P,
  and Sb are 23.7 (22.6) meV/atom, 2.2 (1.4) meV/atom, and 7.4 (1.9) meV/atom,
  respectively. We can see that the strain energy difference from Sb
  ions contributes 74\% to the total strain energy difference. 
  In addition, we find that the difference mainly comes from the 
  deviation of the Al-Sb-Al bond angles from the ideal value (109.47$^{\circ}$).
  In chalcopyrite AlP$_{0.5}$Sb$_{0.5}$ alloy, the maximum
  deviation of the Al-Sb-Al bond angles is 5.4$^{\circ}$, much larger
  than that (2.7$^{\circ}$) in WZ AlP$_{0.5}$Sb$_{0.5}$ alloy.  
  
  The calculated DFT formation enthalpy difference $d\Delta H_f = \Delta H_f (WZA) - \Delta H_f (ZBA)$, 
  where $\Delta H_f ( WZA)$ [$\Delta H_f ( ZBA)$] is the formation
  enthalpy of the WZA (ZBA) 
  defined in Eq.~\ref{eq1}, are shown in Table~\ref{table2}. We see that it follows the
same trend as the strain energy difference, i.e., the GS WZA always has lower formation enthalpy
than the corresponding ZBA. However, the lower formation enthalpy in the WZA does not
  necessarily mean that the WZA has lower total
  energy than the ZBA because the formation enthalpy
  are defined with respect to the pure bulk compounds with the {\it same}
  lattice structure, whereas the total energy difference between the WZ and ZB
  A$_{x}$B$_{1-x}$C alloys should also include the bond energy difference ($dE_b$) 
  between the WZ and ZB phases of the parent binary compounds.
  We define $ E_{WZ-ZB} (AC)$ [$ E_{WZ-ZB} (BC)$] as the energy difference
  between the WZ and ZB phases of the AC (BC) compound. The bond energy
  difference $dE_b(x)$ between the WZA and ZBA as a function of $x$ are then defined as:
  \begin{equation}
    dE_{b}(x) = x E_{WZ-ZB} (AC)  + (1-x) E_{WZ-ZB} (BC).
  \end{equation}
  The total energy difference between the WZA and ZBA  can then be calculated as
  \begin{equation}
    dE_{tot}=d\Delta H_f  + dE_{b}
    \label{etot}
  \end{equation}
  It is clear from Eq. (3) that only when the formation enthalpy difference ($d\Delta H_f$) is more negative
  than $-dE_{b}$, the WZA can be more stable than the ZBA.

  The DFT total energy calculations are performed to determine
  which alloy structure is the GS  phase of
  Ga$_{x}$In$_{1-x}$N, AlP$_{x}$Sb$_{1-x}$,
  ZnS$_{x}$Te$_{1-x}$, and GaP$_{x}$As$_{1-x}$ with $x=0.25$, $0.5$,
  and $0.75$.
  For the parent compounds, we find that the energy differences $E_{WZ-ZB}$ between the WZ and ZB
  phases are $-5.6$, $-10.8$, $3.5$, 
  $6.5$, $3.2$, $6.0$, $8.8$, and $11.4$  meV/atom for GaN,
  InN, AlP, AlSb, ZnS, ZnTe, GaP, and GaAs, respectively.  In
  agreement with previous first principles calculations \cite{Yeh1992} and
  experimental observations, we find that GaN and InN have the WZ GS structure,
  whereas the other compounds take the
  ZB phase as the most stable structure. 
  The DFT results from the alloy calculations are summarized in
  Table~\ref{table2}. For alloys with WZ binary constituents (InN and GaN) or
  small lattice-mismatched ZB binary constituents (GaP and GaAs), the GS
  alloy structure (Ga$_{x}$In$_{1-x}$N and GaP$_{x}$As$_{1-x}$) is the
  same as the parent compounds. However,
  for AlP$_{0.5}$Sb$_{0.5}$, AlP$_{0.75}$Sb$_{0.25}$,
  ZnS$_{0.5}$Te$_{0.5}$,  and ZnS$_{0.75}$Te$_{0.25}$,
  the WZA structure is the GS phase despite that
  the alloys are formed from ZB parent compounds. 
  It is interesting to note that compounds such as MnTe (CdO), which has the stable NiAs (Rocksalt)
  structure can be stabilized in the ZB phase by alloying it with ZB
  compounds \cite{Wei1986,Zhu2008}. 
  Here we show that the alloy can be stabilized in a structure that
  is {\it different from both  parent structures}.
  This remarkable alloy stabilized wurtzite structures
  originate from the fact that the gain in the strain energy
  relaxation when forming the WZA is larger than the average of
  the bond energy difference between the ZB and WZ phases.  
  For example, $d\Delta H_f  = -6.50 $ meV/atom and $dE_{b}=5.01$
  meV/atom for AlP$_{0.5}$Sb$_{0.5}$.
 It is also interesting to see that, the alloy stabilization energy $d\Delta H_f$  
for A$_{0.75}$B$_{0.25}$C is larger than A$_{0.25}$B$_{0.75}$C, i.e.,
  the WZA is more favored
when a large atom is mixed into a smaller host than a smaller atom is mixed into a large host. 
 
\begin{figure}
    \includegraphics[width=7.0cm]{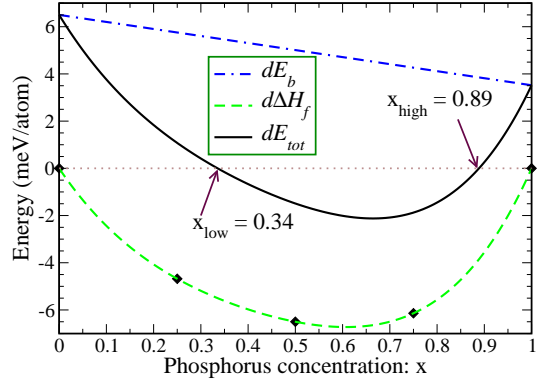}
    \caption{(Color online) Differences in the formation enthalpy ($d
    \Delta H_f$),  total energy ($d E_{tot} $),
    and bond energy ($d E_{b}$) between the WZ and ZB
    AlP$_x$Sb$_{1-x}$ alloys.}
    \label{fig2}
  \end{figure}

  In order to determine the concentration $x$ at which $dE_{tot} <0$, the
  dependence of the difference in the 
  formation enthalpy [$d \Delta H_f(x)$] between the WZA and ZBA
on the concentration $x$ is essential. By definition, 
$d \Delta H_f (0)=0$ and $d \Delta H_f (1)=0$. The $x$ dependence of $d \Delta H_f(x)$ 
can be obtained by fitting the data in Table.~\ref{table2} to a fourth order polynomial.
  The fitted result for the AlP$_x$Sb$_{1-x}$ alloy is shown in
  Fig.~\ref{fig2}. We can see that the curve is asymmetric with respect to $x=0.5$;
the minimum of $d \Delta H_f (x)$ occurs at $x=0.61$. 
  Following Eq.~\ref{etot}, we obtain the dependence of $d E_{tot}$
  on $x$ (Fig.~\ref{fig2}). It is seen that the minimum of
  $dE_{tot}$ occurs at $x=0.66$. And when
  $0.34<x<0.89$, the WZ AlP$_x$Sb$_{1-x}$ alloy is more stable than
  the ZBA. For ZnS$_{x}$Te$_{1-x}$, the result is similar, and the
  lowest concentration and highest concentration for a stable WZ
  ZnS$_{x}$Te$_{1-x}$ alloy are 0.39 and 0.87, respectively.

  In summary, we have identified the GS structures of the A$_x$B$_{1-x}$C WZAs
  with $x = 0.25$, 0.5, and 0.75. Using VFF and DFT calculations, we show
  that the GS WZA always has a lower strain energy and
  formation enthalpy than the
  corresponding ZBA, and thus the strain relaxation favors the
  formation of the WZA. 
  We confirm this idea by showing that GS WZ AlP$_{x}$Sb$_{1-x}$
  (ZnS$_{x}$Te$_{1-x}$) with $0.34<x<0.89$ ($0.39<x<0.87$) is more stable
  than the corresponding ZBA although their parent structures
  crystallize in the ZB phase.

  Work at NREL was supported by the U.S. Department of
  Energy, under Contract No. DE-AC36-08GO28308.
  The work in Fudan (FU) is partially supported by the National
  Sciences Foundation of China, the Basic Research Program of MOE and
  Shanghai, the Special Funds for Major State Basic Research, and
  Postgraduate Innovation Fund of FU.

  \clearpage

  \clearpage

  \clearpage
  \begin{table}
    \caption{VFF-calculated strain energy (in meV/atom) of
      WZ Ga$_{x}$In$_{1-x}$N, AlP$_{x}$Sb$_{1-x}$,
      ZnS$_{x}$Te$_{1-x}$, and GaP$_{x}$As$_{1-x}$ alloys for the $Pmn2_{1}$,
      $P2_{1}$-I, and $P2_{1}$-II structures
      at $x=0.25$ and $0.75$. The numbers in parenthesis are the DFT
      calculated formation enthalpies. $*$ and $\ddag$ indicate the GS
      structures obtained from the VFF and DFT calculations, respectively.}
    \begin{tabular}{cccc}
      \hline
      \hline
      Structures&$Pmn2_{1}$& $P2_{1}$-I &   $P2_{1}$-II \\      
      \hline
      Ga$_{0.25}$In$_{0.75}$N & 12.54$^*$ (13.13) & 14.13 (12.38$^\ddag$) & 14.17 (12.50)\\
      Ga$_{0.75}$In$_{0.25}$N & 20.16 (16.91) & 19.01$^*$ (13.25$^\ddag$) & 19.28
      (13.47) \\
      \hline
      AlP$_{0.25}$Sb$_{0.75}$ & 19.43$^*$ (18.88) & 21.02 (18.19$^\ddag$) & 21.20
      (18.37) \\
      AlP$_{0.75}$Sb$_{0.25}$ & 24.41 (27.09) & 23.46$^*$ (23.35)& 23.69
      (23.32$^\ddag$) \\   
      \hline
      ZnS$_{0.25}$Te$_{0.75}$ & 13.19$^*$ (21.36) & 14.19 (20.76$^\ddag$)& 14.35
      (21.05) \\
      ZnS$_{0.75}$Te$_{0.25}$ & 15.14 (29.36)& 14.73$^*$ (26.96)&
      14.81 (26.86$^\ddag$) \\
      \hline
      GaP$_{0.25}$As$_{0.75}$ & 2.32$^*$ (2.26)  & 2.42 (1.99) & 2.44 (1.96$^\ddag$)\\
      GaP$_{0.75}$As$_{0.25}$ & 2.57 (2.48) & 2.55$^*$ (2.09$^\ddag$)& 2.58 (2.12)\\
      \hline
      \hline
    \end{tabular}
    \label{table1}
  \end{table}
  
  \clearpage
  \begin{table}
    \caption{Differences in the VFF strain energy ($d E_s$), DFT formation
      enthalpy ($d \Delta H_f$), DFT bond energy ($ d E_{b} $), and
      DFT total energy ($d E_{tot}$) between 
      the GS WZAs and ZBAs. Energy is in meV/atom.}
    \begin{tabular}{ccccc}
      \hline
      \hline
      & $d E_s $ & $d \Delta H_f $ &  $ d E_{b} $ & $d E_{tot} $\\
      \hline
      Ga$_{0.25}$In$_{0.75}$N &$-4.03$ &$-5.41$    & $-9.53$     & $-14.94$     \\
      Ga$_{0.5}$In$_{0.5}$N   &$-4.61$ & $-5.79$   & $-8.21$     & $-14.00$      \\
      Ga$_{0.75}$In$_{0.25}$N &$-4.77$ & $-7.25$     & $-6.91$   & $-14.16$       \\
      \hline                                                                
      AlP$_{0.25}$Sb$_{0.75}$ &$-6.18$ &$-4.68$      & $5.76$    & 1.08           \\
      AlP$_{0.5}$Sb$_{0.5}$   &$-7.35$ &$-6.50$     & $5.01$     & $-1.49$       \\
      AlP$_{0.75}$Sb$_{0.25}$ &$-5.52$ &$-6.14$       & $4.27$   & $-1.87$        \\
      \hline                                                                
      ZnS$_{0.25}$Te$_{0.75}$ & $-4.42$&$-3.89$     &5.33        &1.44            \\
      ZnS$_{0.5}$Te$_{0.5}$   & $-5.18$&$-5.43$     &4.62        &$-0.81$         \\
      ZnS$_{0.75}$Te$_{0.25}$ & $-3.67$&$-5.12$      &3.90       &$-1.22$         \\
      \hline                                                                
      GaP$_{0.25}$As$_{0.75}$ & $-0.62$& $-0.78$    &10.75       &9.97            \\
      GaP$_{0.5}$As$_{0.5}$   & $-0.74$ & $-1.05$    &10.12      &9.07           \\
      GaP$_{0.75}$As$_{0.25}$ & $-0.39$ & $-0.85$    &9.48       &8.63            \\
      \hline
      \hline
    \end{tabular}
    \label{table2}
  \end{table}


\begin{thebibliography}{99}
\bibitem{DFT}Our first-principles density functional theory (DFT)
    calculations were performed on the basis of the projector
    augmented wave method \cite{PAW} encoded in the Vienna ab initio
    simulation package \cite{VASP} using the local density
    approximation \cite{LDA}. For relaxed structures, the atomic forces are less
    than 0.01 eV/\AA.
\bibitem{decompose}For each bond involving the atom $i$, half of bond stretching
    energy is contributed to the atom $i$.
    The bond bending energy contributions from atom $i$ include all bond angles 
    which are centered at atom $i$.
\end{thebibliography}
\end{document}